\documentclass[%
 reprint,
 amsmath,amssymb,
 aps,
prb,
]{revtex4-1}

\usepackage{graphicx}
\usepackage{dcolumn}
\usepackage{bm}
\usepackage{colortbl}
\usepackage{color}

\begin{document}

\preprint{APS/123-QED}

\title{Proximity spin-orbit coupling in an armchair carbon nanotube on monolayer bismuthene. }

\author{Marcin Kurpas}
\email{marcin.kurpas@us.edu.pl}
\affiliation{Institute of Physics, University of Silesia in Katowice, 41-500 Chorz\'{o}w, Poland}

\date{\today}

\begin{abstract} 
We study spin-orbit proximity effects in a hybrid heterostructure build of a one-dimensional (1D) armchair carbon nanotube and two-dimensional (2D) buckled monolayer bismuthene. We show, by performing first-principles calculations, that Dirac electrons in the nanotube exhibit large spin-orbit coupling due to a close vicinity of bismuthene. The calculated low-energy band structures of the proximized nanotube display a strong dependence on the position of the nanotube on the substrate, similar to twist-angle dependence found in 2D heterostructures. Based on the first-principles results, we formulate an effective low-energy Hamiltonian of the nanotube and identify key interactions governing the proximity spin-orbit coupling. The proximity-induced spin splitting of Dirac cone bands is in meV range, confirming an efficient transfer of spin-orbit coupling from bismuthene to the nanotube.

\end{abstract}

\maketitle


\section{\label{sec:intro}Introduction}

Two-dimensional (2D) Van der Waals heterostructures have become versatile platforms  to study physical phenomena in low dimensions and promising building blocks for  novel devices for quantum technologies \cite{novoselov2016,ZUTIC201985,Avsar2020}. This was possible thanks to the great ease of modifying their electronic properties, for example, by the proximity effect. It allows a material to acquire new properties by stacking it on top of another material, being the donor of those properties. Graphene, for instance, is essentially free of intrinsic spin-orbit coupling \cite{Gmitra2009,Sichau2019}, but can realize spin Hall states when embedded into a heterostructure with WS$_2$ or MoS$_2$ \cite{Avsar2014,Gmitra_2015, gmitra2016,Safeer2019}.

The effectiveness of the proximity effect to a large extent is dictated by symmetry and physics at the interface between materials.
The presence of a substrate breaks most symmetries of the host material, releasing several constraints on the electron spin and symmetry-allowed spin-orbit terms. Broken space inversion symmetry leads to the emergence of Bychkov-Rashba spin-orbit coupling affecting both, spin splitting and spin texture  of Bloch states \cite{Bychkov1984,Han2014, Gmitra_2015}. Other symmetries, such as in-plane mirror symmetry or pseudospin (sublattice) symmetry can also be broken if present, enabling additional matrix elements of spin-orbit Hamiltonian  \cite{Giovanietti2007,Weeks2011,Kochan2017}. 

The extent to which broken symmetry affects the electronic states depends on the amplitudes of new symmetry-allowed terms. These amplitudes  result from  the interface crystal potential, which reflects mutual interactions and the atomic arrangement of the host material and the substrate. In commensurate 2D heterostructures, the interface potential is lattice periodic, with the period of the supercell. Within the supercell, its shape depends on the misfit and mutual arrangement of materials composing the heterostructure. The latter can be tuned, for example,  by a twist angle, giving  the control over the proximity effect \cite{Li2019,zollner2019,Naimer2021}.

When the host 2D material is replaced  by a one-dimensional (1D) carbon nanotube, the 2D in-plane periodicity of the interface is preserved only in one direction, say $z$ [see Fig. \ref{fig:f1} c)], while in the other [$y$ in Fig. \ref{fig:f1} c)] it is broken due to finite diameter of the nanotube.  It has severe consequences for the electronic states in the nanotube. First, broken rotational symmetry of the nanotube does not protect the longitudinal (along the nanotube axis) component of the electron spin \cite{huertas2006,Klinovaja2011a}. Second, a rapid variation of the crystal potential in the $y$ direction and the non-uniform distance of carbon atoms to the substrate make room for new interface effects absent in flat 2D systems  \cite{Hasegawa2011}.

In this paper we study such effects in  an example hybrid 1D/2D heterostructure made of a (4,4) armchair carbon nanotube and buckled monolayer bismuthene \cite{Nagao2004,Yaginuma2008,Drozdov2014,Akturk2016}. To accurately describe structural changes and interactions between the nanotube and bismuthene  we approach the problem on the atomistic level and perform  first principles calculations based on the density functional theory. 
We show that orbital and spin-orbital properties of the nanotube are very sensitive to the position of the nanotube on the substrate.  By moving the nanotube, we induce qualitative changes in the topology and spin structure of the Dirac cone bands, similar to the twisting of graphene on transition metal dichalcogenides \cite{Naimer2021}.
The band structures topologies  resemble those of armchair carbon nanotubes in coexisting external electric and magnetic fields \cite{Klinovaja2011a}. We use  this resemblance as a hint in  developing an effective low-energy Hamiltonian of the nanotube.

Proximity effects in carbon nanotubes have not been yet intensively studied. So far the focus has been on superconductivity \cite{Morpurgo1999}, charge transfer and contact formation between nanotubes and metal surfaces  \cite{okada2005,Takagi2011,Hasegawa2011,Kuzubov2014,Liao2019}, or  optical properties of the CNT/GeSe heterostructure \cite{Mao2021}.  Carbon nanotubes proximized to a superconductor have also been investigated in the context of topological states \cite{Egger2012,Klinovaja2012,Marganska2018,Milz2019}, but without studying the underlying proximity mechanism. 
Not much is known about spin properties of nanotubes combined with 2D materials or bulk surfaces. This work gives an insight into this topic and demonstrates that hybrid 1D/2D heterostructures are attractive for exploring spin-orbit proximity effects.

 The paper is organized as follows. In the next section we briefly describe methodology and details of first principles calculations. In Section \ref{sec:results} we present and discuss results of the calculations and develop an effective low energy Hamiltonian for the nanotube. Section \ref{sec:conc} contains a summary and closing conclusions.

\section{Methods}
First principles calculations were performed with the plane wave software  package \textsc{Quantum Espresso} \cite{QE-2009,QE-2017}. The heterostructure of CNT and bismuthene was made of 24 Bi atoms building the substrate of width 25.5\,\AA and 48 C atoms forming three unit cells of the (4,4) armchair nanotube. The big width of bismuthene stripe was necessary to eliminate possible bending of bismuthene at  the edges of the simulation cell. 
To ensure the commensurability of the structure, the lattice constant of bismuthene was reduced to $a_{\text{Bi}}=4.4$\,\AA, resulting in its slight compression, of about 3\%, comparing to the experimental value \cite{Monig2005}. The lattice constant of CNT was $a_{\rm cnt}=2.46$\,\AA.  
A vacuum of 16\,\AA was introduced in the $x$ direction (out of plane) to  avoid fictitious interactions between periodic copies of the simulation cell.

\begin{figure}
    \centering
    \includegraphics[width=0.98\columnwidth]{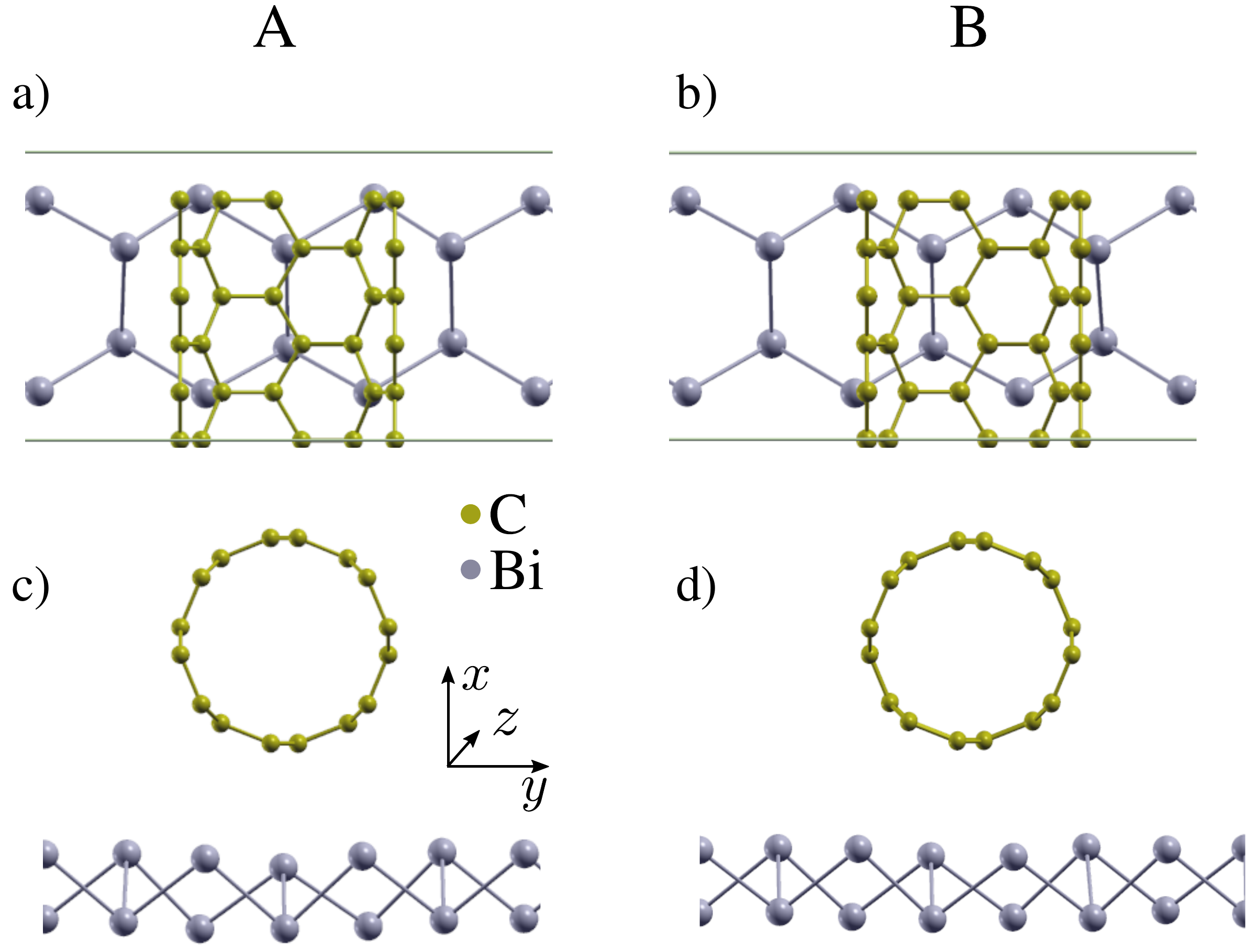}
    \caption{Schematics of the simulated system: a metallic armchair  (4,4) carbon nanotube is placed on monolayer buckled bismuthene. Depending of the position of the nanotube on bismuthene, two different configurations are realized: (a) the zigzag chain of the nanotube is in line with Bi atoms, this is labelled as the A configuration ;(b) the nanotube's bottom zigzag chain is located between two Bi atoms (B configuration). For better visibility, only C atoms closest to the substrate are drawn in (a) and (b). In c) and d) optimized  crystal structures are shown.}
    \label{fig:f1}
\end{figure}

The initial geometry of the heterostructure (Fig. \ref{fig:f1} a), b)) was optimized by minimizing the internal forces acting on atom  using the quasi--Newton scheme, as implemented in {\sc Quantum ESPRESSO}, and assuming ionic minimization convergence criteria: $10^{-3}$\,Ry/bohr for internal forces and  $10^{-5}$\,Ry/bohr for the total energy. During relaxation all atoms were free to move in all directions. 
Taking into account that Bi is a heavy element, we performed optimization independently for the non-relativistic and relativistic calculations, taking the output from the non-relativistic case as the initial structure for the relativistic calculation. Nevertheless, we have not found substantial differences between the two optimized structures. In both cases, the ultrasoft pseudopotentials implementing the Perdew-Burke-Ernzerhof (PBE) \cite{perdew_1996,*perdew_1997}  version of the generalized gradient approximation (GGA) exchange--correlation functional was used, with the kinetic energy cutoff of the plane wave basis sets $48\,$Ry  for the wave function and $485$\,Ry  for charge density. 
Self consistency was achieved with $4\times 10\times 1$ Monkhorst-Pack grid \cite{MPack} while for structure optimization we used a smaller grid  $1 \times 4 \times 1$. All calculations were done including the semiempirical van der Waals corrections \cite{grimme2006,barone2009} and the dipole correction \cite{Bengtsson1999} for a proper determination of the possible energy offset.

For visualization of crystal structures we used the XCrysDen software \cite{KOKALJ1999176}. Fitting of the model Hamiltonian to DFT data was done with help of the least square fitting method implemented in the \textrm{LMFIT} library \cite{lmfit}.

\section{Results and Discussion}
\label{sec:results}
\subsection{First principles results}
We begin with discussing the geometry of the nanotube/bismuthene heterostructure. In Fig. \ref{fig:f1} a),b) we show two configurations studied in this paper. The structures differ in the alignment of the nanotube with respect to the substrate. In configuration A, the lowest zigzag chain of the CNT is in line with the underlying Bi atoms, and C atoms closest to the substrate sit on top of Bi atoms (top position). 
Configuration B is made from A by shifting the nanotube in the $y$ direction by half of the unit cell of bismuthene, such that the lowest zigzag chain of C atoms lies between Bi atoms (hollow position).  

Different initial conditions for A and B setups lead to slightly different responses of bismuthene substrate during structure optimization. In case A, only the atomic Bi chain below the nanotube is pushed down, while in case B, two Bi chains lower their initial position [Fig. \ref{fig:f1} c),d)]. The distances between the nanotube and the underneath Bi atom(s) are 3.28\,\AA~ and 3.15\,\AA~ for the A and B configurations, respectively. We did not notice any substantial changes to the geometry of the nanotube. Energetically the two configurations are very close to each other. The total energy of the whole system for configuration B is 16\,meV lower than for A,  indicating that both are similarly probable for experimental realizations. 

\begin{figure}
    \centering
    \includegraphics[width=0.98\columnwidth]{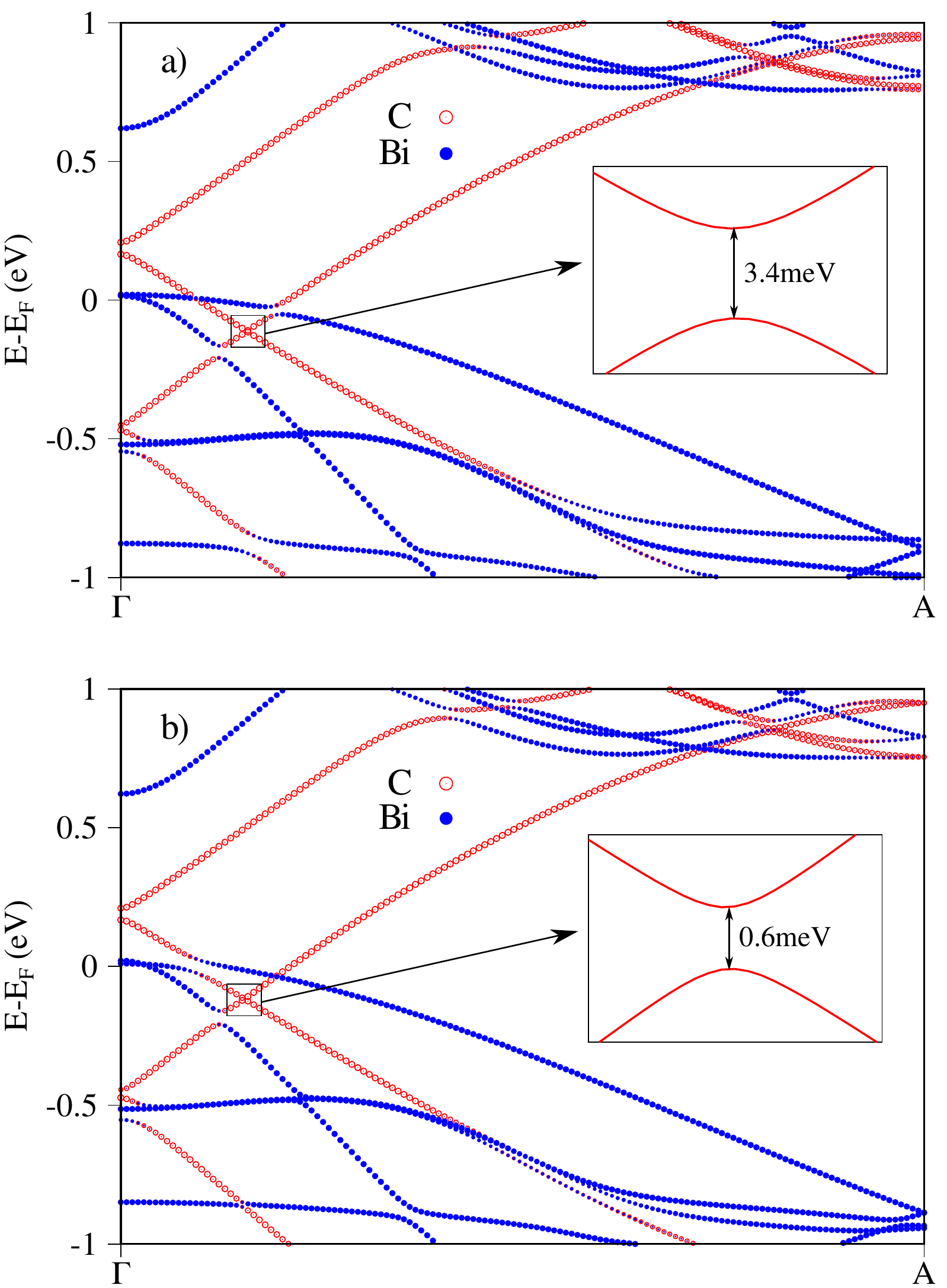}
    \caption{Non-relativistic band structure from first principles calculations. Red color represents C electrons, blue color Bi electrons. a) the band structure for the full First Brillouin zone and A configuration; b) a close view at the region around the Dirac cone of the nanotube on the A configuration; c)same as b) but for B configuration.}
    \label{fig:fig2}
\end{figure}

In Fig. \ref{fig:fig2} a) we show the calculated  non-relativistic band structure for the configuration A plotted along the $\Gamma A$ line in the Brillouin zone of the nanotube. This direction corresponds to crystal momentum  along the nanotube axis parallel to  the armchair edge of bismuthene. 
Bands of the nanotube (red) are easily distinguishable. The Dirac cone lies 0.1\,eV below the Fermi level due to electron doping from the substrate and is well preserved up to $\pm$50\,meV from its center.
Hybridization effects are visible on the left and right sides of the Dirac cone as anticrossings between C and Bi bands. 
A close look at the Dirac cone (inset in  Fig. \ref{fig:fig2} a)) reveals opening a sizeable orbital energy gap $\Delta^\text{A}_{\text{orb}}=3.4$\,meV in the energy spectrum of metallic nanotube due to interaction with the substrate.

The electronic band structure for configuration B is very similar to A (Fig. \ref{fig:fig2} b)). 
Differences are visible in hybridization states at crystal momenta close to the position of the Dirac cone of the nanotube. The anti-crossing at the Fermi energy visible to the right from the Dirac cone for A disappears and is visible to the left from the Dirac cone. A similar effect occurs at energy $\approx -0.75$\,eV below the Fermi level. Most striking is the reduction of the orbital gap, $\Delta^\text{B}_{\text{orb}}=0.6$\,meV, which is  almost six times less than for   case A. 

Since only orbital effects have been considered so far, such significant differences in $\Delta_{\text{orb}}$ between the configurations should  also  be visible in the interface potential. 
\begin{figure}
    \centering
    \includegraphics[width = 0.9\columnwidth]{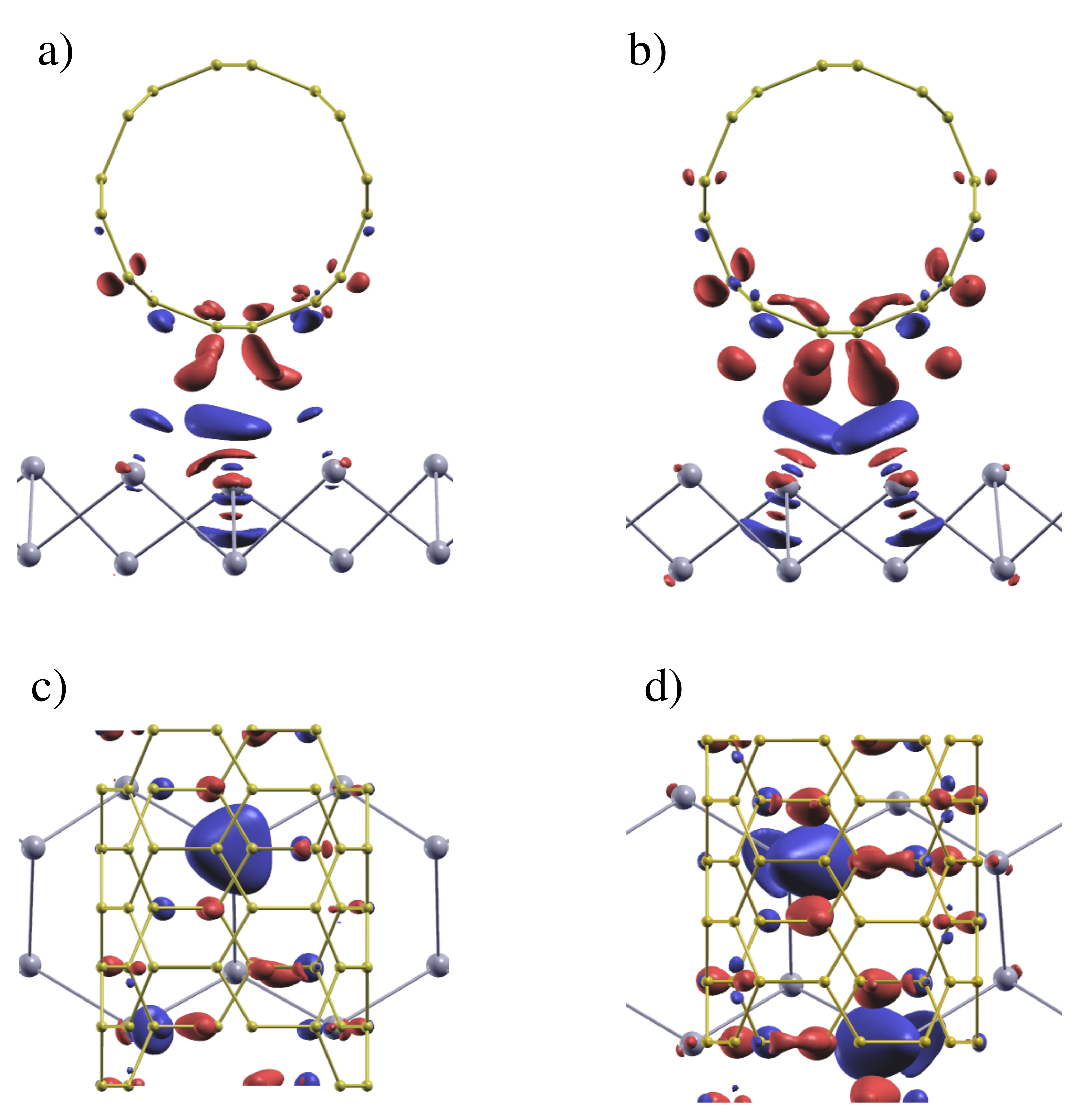}
    \caption{Iso-surfaces of charge density at the interface of CNT and bismuthene within one unit cell: a) side view for the configuration A; b) side view for the configuration B. c), d) top-view for the A and B configurations respectively. The blue (red) color represents negative (positive) values of charge density.}
    \label{fig:cdens}
\end{figure}
Our conjecture is confirmed in Fig. \ref{fig:cdens}, where we show the interface charge density iso-surfaces $\rho(\bf{r})_{\text{int}}$,
\begin{equation}\label{eq:rho}
    \rho(\bf{r})_{\text{int}} =  \rho(\bf{r}) - \rho(\bf{r})_{\text{Bi}} - \rho(\bf{r})_{\text{cnt}}.
\end{equation}
Here $\rho(\bf{r}) $ is the total charge density of the full system, and $\rho(\bf{r})_{\text{Bi/cnt}}$ is the charge density of the slab/nanotube respectively. 
Qualitative differences in $\rho(\bf{r})_{\text{int}}$  between A and B configurations are evident. For the former, a single big charge pocket (big blue lob in Fig. \ref{fig:cdens}c)) is formed (approximately) below one C atom,  while for the latter two lobs are localized below C atoms belonging to different sublattices of the nanotube (Fig. \ref{fig:cdens}d)). 
As we will show below, details of the interface potential also strongly impact the spin-orbit coupling in the nanotube.

Relativistic effects in carbon nanotubes are much stronger than in flat graphene due to curvature-induced hopping between $\sigma$ and $\pi$ orbitals \cite{Ando2000,huertas2006,Kuemmeth2008,jhang2010,Steele2013}.
In armchair nanotubes, effects of intrinsic spin-orbit coupling in the band structure are manifested by opening a spin-orbital gap a the $K$-point, while bands remain spin doublets, which is guaranteed by space inversion and time reversal symmetry.  Theory predicts, that the value of the spin-orbital gap at the Dirac point is $\Delta_{\text{so}}^\text{K}=0.6$\,meV/$d$ -- 0.85\,meV/$d$ , where $d$ is the diameter of the nanotube in nanometers  \cite{Izumida2009,zhou2009}. This is roughly two orders larger than for graphene, for which $\Delta_{\text{so}}^\text{K}\approx 25$\,$\mu$eV~--~40\,$\mu$eV \cite{Gmitra2009,Sichau2019}.
Our first principles calculations give for pristine (4,4) nanotube of  diameter $d=0.55$\,nm, $\Delta_{\text{so}}^\text{K}=1.55$\,meV, in a good agreement with the above formula.

\begin{figure}
    \centering
     \includegraphics[width=0.98\columnwidth]{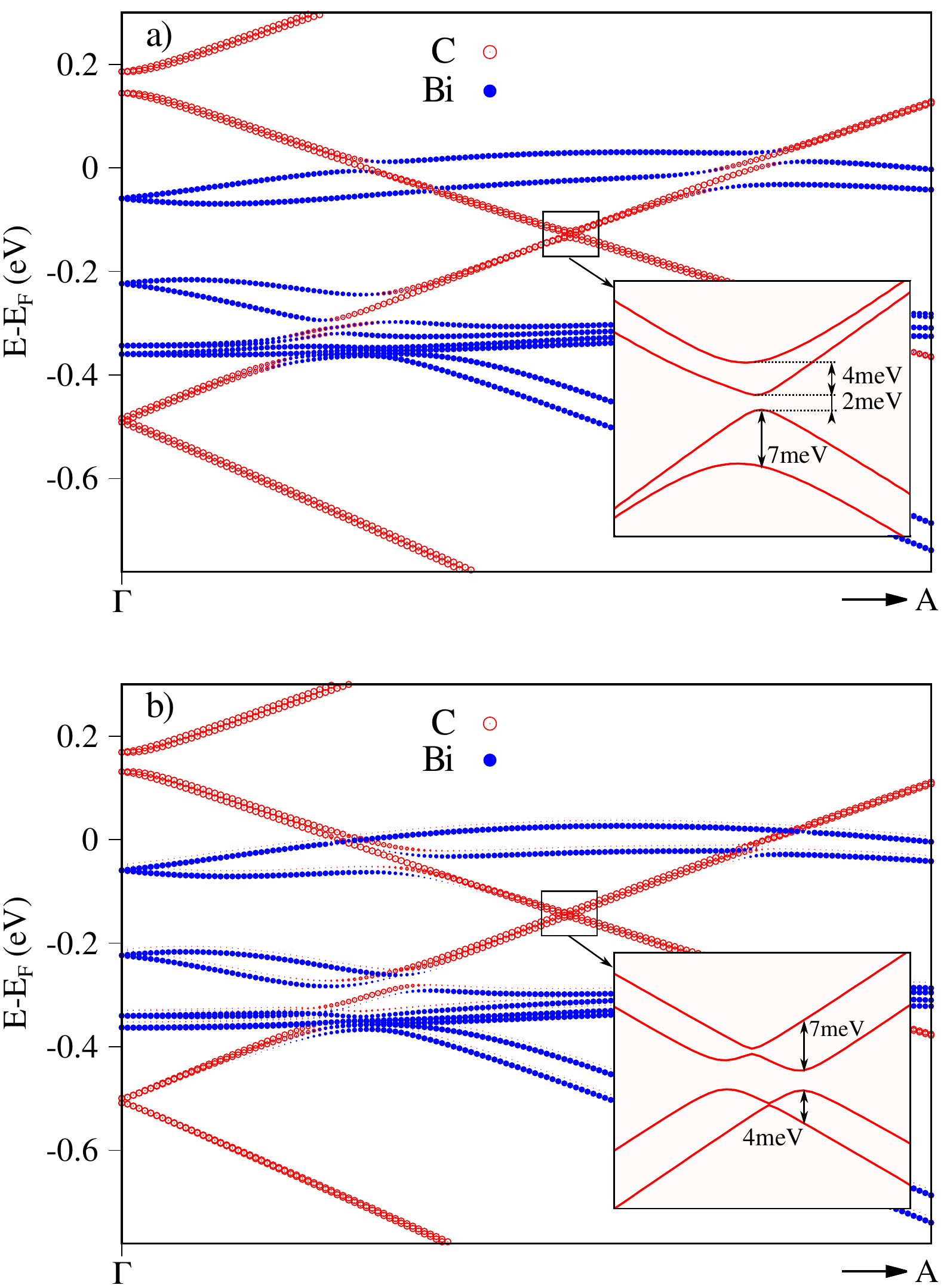}
    \caption{Relativistic band structure from first principles projected onto C (red) and Bi (blue) atomic states. Panels a) and b) correspond to the A and B configurations, respectively. Strong repulsion and spin splitting in Bi bands modifies hybridization between states of CNT and bismuthene compared to the non-relativistic calculation.  }
    \label{fig:fig_rela}
\end{figure}

In Fig. \ref{fig:fig_rela} a), b) we show calculated relativistic band structures of CNT/Bi heterostructure projected onto the atomic states of Bi and C. The inclusion of SOC removes the orbital degeneracy of Bi states at the Fermi level pulling  them apart by 164\,meV. Spin states are also split off  by energy of about 50\,meV, which is unsurprising considering the sizeable atomic number of bismuth Z=83 and strong spin-orbit coupling in bismuthene \cite{Kurpas2019}.

Similar to the non-relativistic case, the Dirac cone of the nanotube is well separated from Bi bands. A closer look  reveals differences in the band structure topology for  A and B configurations.
For the former, the top valence and bottom conduction bands of the nanotube  meat at the same $k$-point but the outer spin subbands are misaligned, leading to a pronounced asymmetry of spin splitting $\Delta_{\uparrow\downarrow}$ between the right and left movers (see inset in Fig.  \ref{fig:fig_rela} a)). 
The  splitting is also particle-hole asymmetric. In the hole branch states at the band maximum are split off by $\Delta_{\uparrow \downarrow}=7$\,meV, while at the conduction band minimum $\Delta_{\uparrow\downarrow}=4$\,meV. In both cases the values largely exceed splittings induced by a transverse external electric field. Our first principles calculations give splitting energy on the order $\approx 10$\,$\mu $eV/Vnm$^{-1}$, on the same level as for graphene \cite{Gmitra2009}.

The corresponding spin expectation values of Dirac cone bands are shown in Fig. \ref{fig:model_pos1} b)-d). Besides the $S_y$ component, expected from the crystal potential gradient in the $x$ direction (normal to the surface of bismuthene), also $S_x$ is pronounced. $S_z$ component in weak, which can be understood a consequence of the lack of an intrinsic Zeeman field polarizing spins along the tube axis in chiral and zigzag tubes \cite{Izumida2009}.

A qualitative different picture of the Dirac cone bands is seen for configuration B [see inset in Fig. \ref{fig:fig_rela} b)]. Instead of a vertical energy splitting of spin states we observe a horizontal shift of bands in crystal momentum, in opposite direction for each band, resulting in the emergence of two cones of similar energies but opposite spin. This is confirmed by tracing the spin texture of bands shown in Fig. \ref{fig:model_pos2} b)-d). For instance, the spin expectation in the band VB1 changes from 0.5 to -0.5 at the band maximum ($k\approx -0.008$\,nm$^{-1}$),  but at $k=0$ the two cones interchange and spin turns back towards 0.5. 
In contrast to case A, in configuration B, the gap between the branches significantly increasesfrom 0.6\,meV in the non-relativistic case to 2.4\,meV, indicating its spin-orbital origin. 

\subsection{Effective Hamiltonian} 
The topology of Dirac cone bands is very similar to the energy spectrum of armchair nanotubes in coexisting transverse electric and magnetic fields \cite{DeGottardi2009,Klinovaja2011a}. 
We exploit these similarities to build an effective model capturing essential features shown by first principles results. 
We start from the orbital Hamiltonian of an  infinite pristine armchair nanotube describing linear dispersion of the Dirac cone \cite{Klinovaja2011a}
\begin{equation}
  H^{arm}  = \tau \hbar v_F k \sigma_2.
  \label{eq:h_klin}
\end{equation}
Here, $\tau=\pm 1$ is the valley index, $v_F$ is the Fermi velocity, $k$ is longitudinal crystal momentum, and $\sigma_2$ is the Pauli matrix acting on the sublattice degree of freedom.
To describe opening of the orbital gap we introduce a staggered on-site potential $\Delta_{st}$ diagonal in spin basis
\begin{eqnarray}
H^{st} = \Delta_{st} \sigma_3.
\label{eq:H_mass}
\end{eqnarray}
The parameter $\Delta_{st}$ includes the contribution from  several factors affecting the orbital gap, such as, the deformation of the nanotube and renormalization of the gap by crystal potential from the substrate.  
We use the orbital Hamiltonian $H^{orb} = H^{arm}+ H^{st}$  to find initial values for $v_F$ and $\Delta_{st}$ by fitting the model to the first principles energy spectrum. 

Next, we add effects of  spin-orbit coupling. The intrinsic spin-orbit coupling opening a spin-orbital gap in at the $K$-point in pristine armchair nanotube is given by \cite{Izumida2009, Klinovaja2011a} 
\begin{equation}
    H^{int} = \alpha S_z \sigma_1,
\end{equation}
where  $S_z$ is the spin one-half operator with eigenvalues $\pm 1$, and $\alpha$ is a parameter defining the strength of SOC.

The interaction of the nanotube with the substrate changes the charge density profile and generates non-vanishing crystal potential gradients, or equivalently, electric fields. A static electric field  in the $x$ (stacking)  direction generates the Hamiltonian \cite{Klinovaja2011a}
\begin{equation}
    \label{eq:Hef}
    H^R = \tau E_x S_y \sigma_2,
\end{equation}
where  $E_x$ is the electric field strength in meV.
This term is responsible for symmetrically splitting spin subbands in each branch of the Dirac cone and spin polarization in the $y$ direction.

Finally, we define the time reversal-symmetric effective spin-orbit Hamiltonian $H^{so}$ that, along with Eq. (\ref{eq:Hef}), will play the central role in reconstructing the first principles band structure of the nanotube
\begin{equation}
   H^{so} = \tau \Omega_{1} S_y\sigma_0 + \tau \Omega_2 S_y\sigma_3  + \Omega_3 S_y \sigma_1 + \tau \Omega_4 S_z\sigma_2.
    \label{eq:zeeman}
\end{equation}
The parameters $\Omega_{i}$, $i=\lbrace 1,2,3,4\rbrace$ are amplitudes of effective spin-orbit fields. 
Since all symmetries of the nanotube are broken, there are no constraints on the form of terms appearing in $H^{so}$. Thus, any term of the product $(\bm{p}\times \nabla V(\bm{r}))\cdot \bm{\sigma}$, including $H^R$, is allowed, provided it does not break time reversal symmetry.
However, we kept minimal number of possible terms allowing us to reconstruct the first principles energy spectrum and the spin texture. 

The first term in (\ref{eq:zeeman}) can be interpreted as a Hamiltonian of an effective magnetic field in the $y$ direction, perpendicular to the electric field $E_x$ and to the nanotube axis ($z$ axis).  It is sublattice even, and is similar to the  Hamiltonian of the intrinsic chiral magnetic field in zigzag and chiral nanotubes, $H_Z = \tau \beta S_z \sigma_0$, where the parameter $\beta$  depends of the chiral angle $\theta$,  $\beta \sim 3\theta$ \cite{Izumida2009,Jeong2009}. For armchair carbon nanotubes $\theta=\pi/6$ giving $\beta=0$.
Here, the field polarizes spins in the direction transverse to the tube axis, thus its origin is of different nature that of  $H_Z$. 
Besides modifying the energy of spin states, already split off by the electric field $E_x$, it introduces a spin--dependent shift in $k$ leading to an asymmetry of spin splitting for left and right movers.

The second term in (\ref{eq:zeeman}) describes a staggered effective magnetic field with opposite sign on the A and B sublattices (sublattice odd).  As we will show, these two terms in (\ref{eq:zeeman}), together with $H^R$,  play a dominant role in reconstructing energy spectrum and main features of the spin texture. The remaining two terms in (\ref{eq:zeeman})  are required for correct reconstruction of spin expectation values. They involve sublattice mixing via operators $\sigma_1$ and $\sigma_2$, in contrast to $\Omega_1$, and $\Omega_2$, which act on a given sublattice.

\begin{figure}[h]
    \centering
    \includegraphics[width=0.99\columnwidth]
    {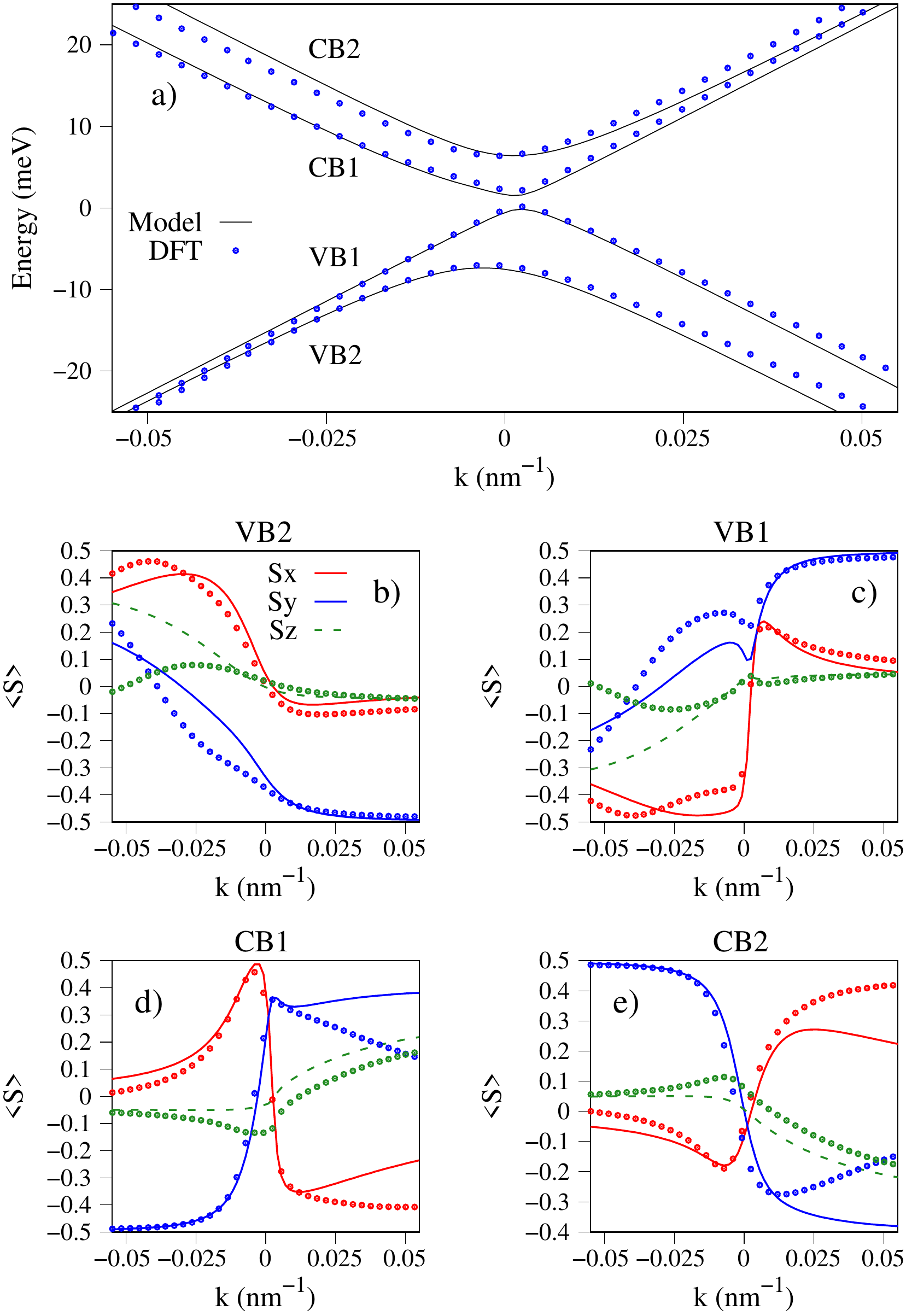}
    \caption{Electronic properties of (4,4) armchair carbon nanotube on monolayer bismuthene for the configuration A. a) Dirac cone bands of the nanotube; b)-d) corresponding spin expectation values of bands shown in a). Dots represent first principles data while lines are results from the effective Hamiltonian  (\ref{eq:full_model}).}
    \label{fig:model_pos1}
\end{figure}

\begin{figure}
    \centering
    \includegraphics[width=0.99\columnwidth]
    {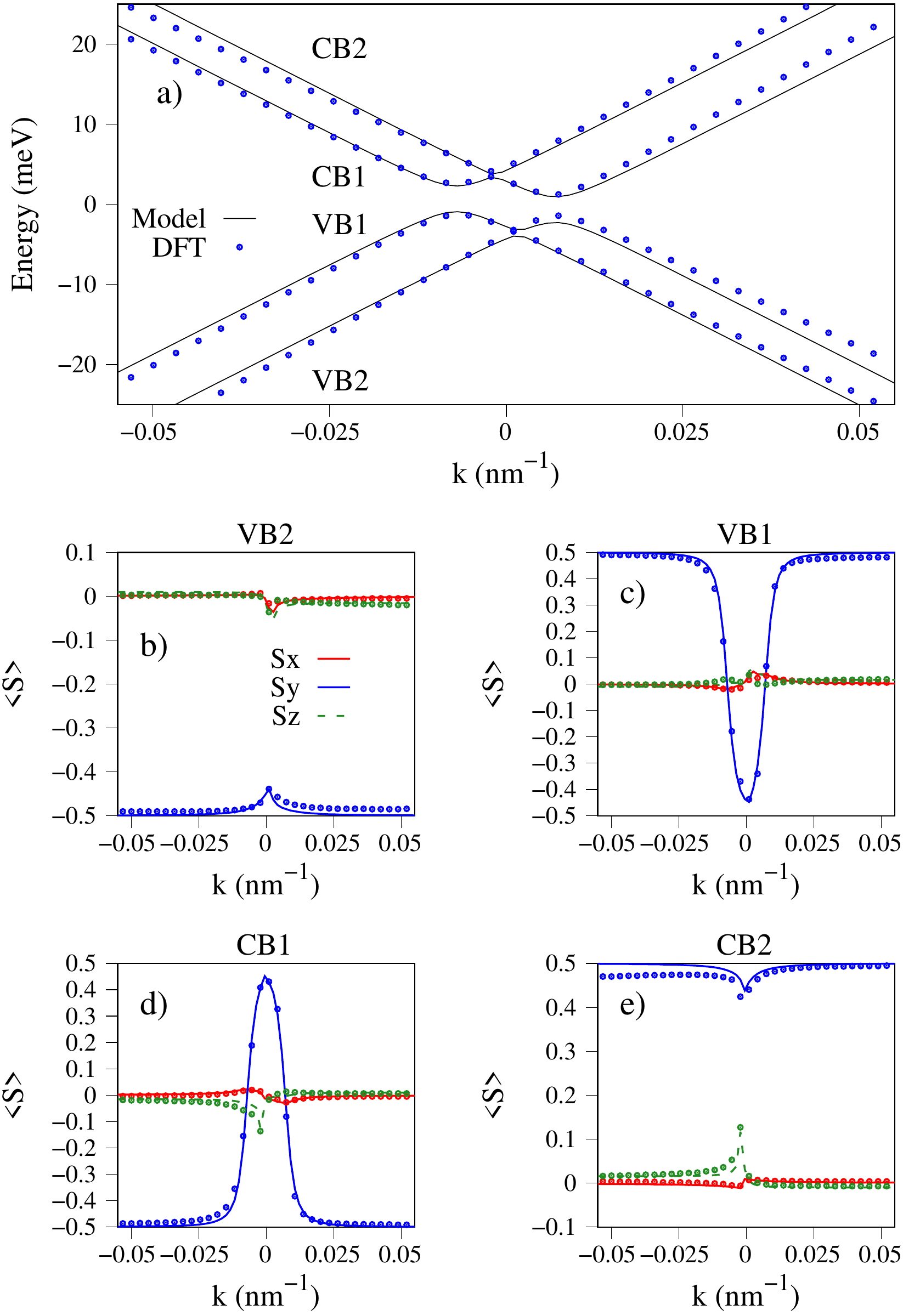}
    \caption{Same as Fig. \ref{fig:model_pos1} but for the configuration B.}
    \label{fig:model_pos2}
\end{figure}

The parameters of the effective model are found by least squares fitting the full Hamiltoanian 
\begin{equation}
   H=H^{orb}+H^{int}+H^R+H^{so}
   \label{eq:full_model}
\end{equation}
 to DFT data. Theoretical results superimposed on the first principles data are shown  in Figs. \ref{fig:model_pos1} and \ref{fig:model_pos2}, while the corresponding parameters are collected in Table \ref{tab:model_parameters}. 
 
 The overall agreement of the model with DFT data is very good. Some discrepancies in spin texture are visible in Fig. \ref{fig:model_pos1} a) - d), showing  a rather complicated nature of the induced spin-orbit coupling in this case, probably by the nearby hybridization with the Bi bands. 
Looking at the values listed in Table \ref{tab:model_parameters} it is clear that the configurations A and B  activate different orbital and spin-orbital fields generated by the interface crystal potential. While for the configuration A, terms with $\Delta_{st}$, $E_x$, $\Omega_1$ and $\Omega_2$ are essential for reconstruction of the band structure and the spin texture, in the case B, the terms with $E_x$, and $\Omega_1$ are sufficient to get a good agreement with DFT data 
The significant value of $\Omega_1$ in the latter case demonstrates the dominant contribution from the Zeeman-like field, which here is of extrinsic origin.

The topologies of Dirac cone bands of the nanotube  discussed above resemble those of graphene on transition metal dichalcogenides, where the spin-orbital proximity effect was controlled by the twist angle between the components of the heterobilayers \cite{Li2019,david2019,Naimer2021}. Here, a similar effect is achieved by introducing a shift instead of a twist, which was shown to have marginal effects on proximity spin-orbit coupling in 2D heterostructures \cite{gmitra2016}.

\begin{table*}
\caption{\label{tab:model_parameters}
Parameters of the effective Hamiltonian (\ref{eq:h_klin}) obtained  by least square fitting to the first principles data.}
\begin{ruledtabular}
\begin{tabular}{ccccccccc}
      Configuration & $\Delta_{st}$ (meV) & $v_F$ (nm$\cdot$ s$^{-1}$) & $ E_{x}$ (meV)& $\Omega_1$ (meV)&$\Omega_2$ (meV)& $\Omega_3$ (meV)& $\Omega_4$ (meV) & $\alpha$ (meV)\\ \hline
        A & 1.74 & $6.9\cdot 10^{14}$&-1.75 &  1.4  & -2.6 & 0& -0.3 & 3.2\\
       B & 0.17 &$6.8\cdot 10^{14}$ & 0.67  &  3.2& -0.16 & 0.1  & -0.07& 1.6 \\

\end{tabular}
\end{ruledtabular}
\end{table*}

\section{Conclusions}\label{sec:conc}
We have studied proximity spin-orbit coupling in a hybrid 1D/2D heterostructure made of an armchair carbon nanotube and buckled monolayer bismuthene. Using first-principles calculations, we have found that Dirac electrons of the nanotube are very sensitive to the interface crystal potential reflecting the atomic environment created by the substrate. We modified the latter by changing the position of the nanotube and induced qualitative changes in the topology and spin texture of Dirac cone Bloch states. These changes were triggered by a change of effective spin-orbital fields at the interface  being the analogs of external magnetic and electric fields, as we showed by an effective model analysis.
The proximity-induced spin-orbit coupling in Dirac cone is on meV range, much beyond the capabilities of a transverse external electric field, which demonstrates an effective transfer of spin-orbit coupling from bismuthene to the nanotube.

\begin{acknowledgments}
The authors thanks M. Marganska-Lyzniak,  M. Gmitra, M. Milivojevi\'c, and J. Fabian for fruitful discussions. The authors acknowledge support from the Interdisciplinary Centre for Mathematical and Computational Modelling (ICM), University of Warsaw (UW), within grant no. GA84-43. The project is co-financed  by the National Center for Research and Development (NCBR) under the V4-Japan project BGapEng V4-JAPAN/2/46/BGapEng/2022.
\end{acknowledgments}

\bibliography{bibliography}

\end{document}